# Large Scale Synthesis of Bi-layer Graphene in Strongly Coupled Stacking Order

By *Zhiqiang Luo, Ting Yu, Jingzhi Shang, Yingying Wang, Sanhua Lim, Lei Liu, Gagik G. Gurzadyan, Zexiang Shen\*, and Jianyi Lin\**

[*] Prof. Z.X. Shen, Z.Q. Luo, Prof. T. Yu, J. Z. Shang, Y. Y. Wang, L. Liu, Prof. G. G. Gurzadyan

Division of Physics and Applied Physics, School of Physical and Mathematical Sciences, Nanyang Technological University, Singapore, 637371

E-mail: Zexiang@ntu.edu.sg;

[*] Prof. J. Y. Lin, S. H. Lim

Applied Catalysis, Institute of Chemical and Engineering Sciences, Singapore, 627833

E-mail: lin_jianyi@ices.a-star.edu.sg

Large scale synthesis of single layer graphene (SLG) by chemical vapor deposition (CVD) has received a lot of attention recently. However, CVD synthesis of AB stacked bi-layer graphene (BLG) is still a challenging work. Here we report synthesis of BLG homogeneously in large area by thermal CVD. The 2D Raman band of CVD BLG splits into four components, suggesting splitting of electronic bands due to strong interlayer coupling. The splitting of electronic bands in CVD BLG is further evidenced by the study of near infrared (NIR) absorption and carrier dynamics probed by transient absorption spectroscopy. Ultraviolet photoelectron spectroscopy invesigation also indiates CVD BLG possesses different electronic structures from those of CVD SLG. The growth mechanism of BLG is found to be related to catalystic activity of copper (Cu) surface, which is determined by purity of Cu foils employed in CVD process. Our work shows that strongly coupled or even AB stacked BLG can be grown on Cu foils in large scale, which is of particular importance for device applications based on their split electronic bands.

# 1. Introduction

Graphene, a single layer of carbon atoms arranged in a honeycomb lattice, has attracted great attentions because of its novel physical properties, such as anomalous quantum hall effect and ballistic transport of massless Dirac fermions, which are closely related to unique linear energy-momentum dispersion in its electronic bands.[1] In addition to SLG, BLG, a fancinating system of two stacked SLG, demostrates tunable band stuctures with different staking orders, i.e. rotation angles between two graphene layers, both predicted theoretically and observed experimentally.[2-8] As the most energetically preferred commensurate stacking order, BLG in AB (Bernal) stacking, a stack of two SLG in such a way that a carbon atom in one layer is in the center of the other layer's hexagon (rotation angle defined as 60°), has a unit cell of four carbon atoms accompanied by strong interlayer coupling, causing the π electron linear dispersion splits into two parabolic branches near K point.[5] BLG in AA stacking with a rotation angle of 0° also shows strong interlayer coupling, which makes split linear energy bands intersected with each other and the K point at Fermi energy changed to two neighboring points.[5]

Regarding BLG with other rotation angles in a range of 0~60°, their band stractures renormalize in presence of a larger unit cell and weaker interlayer coupling strength than those of the AB and AA stacked BLG.[4] Recent calculation shows that there is a critical angle around 1.5° and a symmetric angle at 30°.[2-4] For rotation angles below 1°, parabolic band dispersion of BLG is obtained due to strong interlayer coupling, while for angles above 2° up to the symmetric point of 30°, BLG becomes weakly coupled or decoupled. For such BLG, their dispersion near the K ponit is linear the same as SLG, but modified with shorten linear dispersion range, depressed fermi velocity, as well as reduced energy span between conduction band and valence band at M point.[2-4] The stacking order and corresponding interlayer coupling strength dependent electronic properties of BLG offer demandable physical properties for device applications. For example, the AB staked BLG shows a remarkable ability to tune band gap in electronic band structures and electron-phonon interactions by a perpendicular electric field effect,[8,9] which can be developed to be a new kind of phonon laser. When the graphene layers in BLG rotated with a samll angle within 1~2°, low-energy Van Hove singularity in density of states was theoretically predicted and experimetally observed by scanning tunneling spectroscopy (STS) measurements, implying such BLG should be new materials with superconductivity and charge density wave phases.[3,10]

It is of particular interest to explore growth of BLG in large scale with controlled staking orders for practical applications based on their predesinged properties. Up to now, large scaled graphene layers, which can be fabricated by epitaxial growth on SiC substrate as well as CVD on metal substrates, always shows obvious decoupling of the graphene layers.[11,12] For instance, both transport measurement and angle-resolved photoemission spectroscopy (ARPES) investigation indicate the multi-layer epitaxial graphene grown on the carbon terminated SiC surface preserve the linear dispersion as the SLG does.[11] CVD graphene layers grown on nickel (Ni) films also show decoupling or weak interlayer coupling as indicated by single Lorentzian lineshape of double resonance 2D Raman band.[12] In this article, we report synthesis of strongly coupled BLG in large area. The splitting of the electronic bands caused by strong interlayer coupling is probed by the Raman spectroscopy, in particular by monitoring the splitting of 2D Raman band. The growth mechanism of such BLG system is found to be related to catalystic activity of Cu surface, which is determined by the purity of Cu foils. The influence of interlayer coupling on electronic structrues and carrier dynamics of as-grown BLG are invesitigated in comparison to those of CVD SLG by ultraviolet photoelectron spectroscopy (UPS), optical absorption spectroscopy and transient absorption spectroscopy.

## 2. Results and Discussion

### 2.1 CVD Synthesis and Raman spectrum of BLG in strongly coupled stacking order

Large area synthesis of high quality SLG on Cu substrates by CVD method has been well developed by several research groups.[13-15] Cu foils of 99.95% purity (Goodfellow, UK) was used in our CVD process for graphene synthesis. With growth parameters described in experimental section, it took 30 min for growth of a SLG on the Cu surface. While interestingly, we noticed a second graphene layer can be grown on the first graphene layer in additional 180 min CVD process. To determine the number of graphene layers, a simple yet efficient optical approach, contrast spectroscopy of graphene on a $SiO_2$/Si substrate was employed.[16-18] Figure 1a is an optical image of the CVD graphene layers transferred onto a 285 nm $SiO_2$/Si substrate, which shows clear contrast difference for several regions in the graphene sheet, indicating different thicknesses or number of layers. The contrast spectra taken from three representative areas, A, B and C of potentially having three different thicknesses, are shown in Figure 1b. The maxium contrast value locating around 560 nm explains the visibility of the graphene layers to naked eyes.[16] The peak value of the contrast spectra of the A, B and C regions are 0.11, 0.18, and 0.27, respectively, close to the calculation values of those of SLG, BLG and three layer graphene

(TLG).[16] After intensive investigations of the optical images and their corresponding contrast spectra, it is found that 90% area of the as-grown graphene sheet is BLG. The rest 10% is mainly TLG with very small fractions of SLG and multilayer graphene (MLG).

Raman spectroscopy/imaging has been widely applied to exploit the structural and electronic properties of graphene, including number of layers, stacking order, edge states, strain effect, band structures and doping concentration.[19-26] As shown in Figure 2a, Raman spectra of graphene layers display two characteristic peaks at ~1580 cm$^{-1}$ and at ~2670 cm$^{-1}$, corresponding to G band, in-plane vibration mode ($E_{2g}$ phonon at Brillouin zone center), and 2D band, inter-valley double resonance scattering of two TO phonons around the K point of Brillouin zone, respectively.[19] A variety of Raman mode features such as intensity of G mode and width of 2D appears in spectra recorded from different parts of the as-grown graphene sheet. To exploit the details, Raman images are generated by extracting the Raman intensity of G and 2D bands, the peak position and full width at half-maximum (FWHM) of 2D band, as displayed in Figure 3a, 3b, 3c and 3d, respectively. The dark (smaller G band intensity) area in Figure 3a, labeled as area A is a small domain of SLG identified by a typical Raman spectrum of SLG (see Figure 2a).[19,20] The medium contrast (appearing in red) area labeled as area B in Figure 3a dominates the entire as-grown graphene sheet. The corresponding Raman spectrum of this area exhibits comparable intensity of G and 2D bands together with a relatively broader asymmetric 2D peak, a standard Raman feature of BLG.[19,20] Two small bright (in yellow) areas in Figure 3a and 3b are noticed and labeled as area D and E. The spectrum of the former shows a strong and sharp 2D band, which is analogous to that of SLG but with an apparent blue shift (13 cm$^{-1}$) caused by the reduction of Fermi velocity, the same as Raman spectrum of the decoupled or weakly coupled BLG previously demostrated in CVD graphene layers grown on Ni substrates and the mechanically cleaved graphene layers in folded configuration.[6,12] The spectrum of the latter has a super strong G band with a $I_G/I_{2D}$ ratio of more than 10, which results from G band double resonance Raman scattering in BLG with special rotation angles as demonstrated in our previous work.[7] The D and E type areas are widely distributed in CVD BLG sheet but in very small fractions.

According to double resonance theory, the 2D Raman band reflects electronic structures of graphene layers.[19] Figure 2b shows the Raman spectra of the interested areas discussed above in 2D band range. In contrast to the spectra of area D and E, which present a single Lorentzian lineshape 2D band, the spectrum of area B displays an asymmetry lineshape of 2D band, which cannot be fitted as one Lorentzian peak but multipile components. It is noticed that in some areas of the B region, for example C spot indicated in Figure 3c, the 2D bands can be well fitted by four components individually with FWHM of 30~35 cm$^{-1}$. The intensity of the lowest and the highest frequency sub-bands is weaker, which is well analogous to the split 2D band of AB stacked BLG.[19] It has been interpreted theoretically that the splitting of 2D peak in AB stacked BLG originates from the four double-resonance Raman processes, a result of the splitting of electrons dispersion in valence and conduction bands into two parabolic branches cuased by strong interlayer coupling.[19] Therefore, it is unambiguous that most of the areas in our synthesized CVD BLG exhibit strong interlayer coupling though they might deivate a bit from exact AB stacked BLG due to small rotation angles and/or larger interlayer separation. We believe perfect AB stacked BLG could be fabricated by CVD method with further effort.

In addition to the Cu foil with purity of 99.95%, Cu foils with purity of 99.999% and 99.8% (Alfa Aesar, US) were also employed in our CVD process for comparison. It is found that after growth of SLG on the 99.999% Cu foil within 30 min, the second layer cannot be grown even with a longer additional CVD duration, i.e. more than 180 min. The Raman spectrum (see Figure 4) of the graphene grown on the 99.999% Cu at 1000 °C for 210 min shows typical characteristic features of SLG. This is similar to previous report on growth of predominant SLG in pyrolysis of $CH_4$ and $H_2$ on Cu foils reported by Li et al.[13] The homogeneous and large area growth of BLG in our CVD process should benifit from the lower purity (99.95%) of the Cu foil, where the dopants or impurities could effectively enhance catalytic performance of Cu surface as previuosly proved in our growth of carbon nanotubes/nanofibers on Cu catalyst.[27] A direct evidence of higher activity of the 99.95% Cu foils is that SLG flims can be grown at a temperature as low as 800°C (see its Raman spectrum in Figure 4). On the contrary, no continous graphene films can be formed on the surface of 99.999% Cu foil at such a low temperature in 30 min. A CVD process on 99.8% Cu foils at 1000°C for 30 min leads to the formation of graphene films with large fractions of MLG in a background of SLG and BLG. The fraction and the domain size of the MLG can be significantly enlarged by prolonging growth durations. At 1000°C, catalytic activity of the 99.8% Cu is much higher compared to that of a more purified Cu. Kinetic factors such as the surface reaction rate plays a critical role on the uniformity of thickness of CVD graphene layers by limiting the deposition of carbon atoms on Cu surface. Faster surface reaction rate, for example on the 99.8% Cu face, results in loss of thickness uniformity, which is consistent with recent kinetic study of CVD graphene on Cu foils.[28] Thus, it is clear that appropriate catalytic activity of Cu foils and careful controlling of CVD kinetics are important for homogeneous growth of BLG.

## 2.2 Electronic structures of strongly coupled BLG

The influence of interlayer coupling on electronic structures of CVD BLG was investigated in comparison of its valence bands probed by UPS with those of CVD SLG and highly ordered pyrolytic graphite (HOPG). As shown in Figure 5, all the He II valence band spectra show five characteristic states of graphite, which are assigned to: (1) C 2p π between 0 and 4 eV; (2) the crossing of C 2p π and C 2p σ bands around 5.9 eV (2p π+ σ) ; (3) C 2p σ at 7.9 eV; (4) C 2s-2p hybridized state at 10.6 eV; and (5) C 2s σ band at ~13.4 eV.[29] With the increase of graphene layer numbers, there are two apparent differences among the spectra, (1) narrowing of the FWHM for the 2p π and 2p π+ σ states, and (2) blue shift of 2p π state and red shift of 2s-2p hybridized state in CVD SLG. The FWHM of the 2p π state for the CVD SLG, CVD BLG and HOPG are 1.45 eV, 1.2 eV and 1 eV, respectively. Similar to previous UPS study of multi-walled carbon nanotubes and graphite, the origin of UPS bandwidth for graphene should be mainly attributed to photohole life time broadening and phonon broadening.[30,31] Generally, these two broadening mechanism are controlled by Auger quenching process and hole-phonon coupling, respectively, both of which are determined by electronic band structures and corresponding density of the states.[32] Therefore, the FWHM of CVD BLG in between those of CVD SLG and HOPG indicates that electronic structures of CVD BLG should deviate from those of CVD SLG and approach those of HOPG, which consist of predominant fractions of AB stacked graphene layers.[33] For better understanding of the electronic structures of our CVD BLG, especially the details in splitting of electron bands due to interlayer interaction, ARPES characterization is required to be performed in future.[34] Regarding to the binding energy shift of the state peaks in CVD SLG, it should result from the curvature of ripple structures in SLG and corresponding rehybridization of π and σ states, which was previously revealed in multi-walled carbon nanotubes by our UPS investigation.[31] While there are less amount of ripples in BLG, resulting in little influence on both atomic and electronic structures of BLG.[35]

The electronic structures of CVD BLG was further investigated with optical absorption spectra in UV-Vis-NIR range. Figure 6 displays the absorption spectra of both CVD SLG and CVD BLG for comparison. The absorption spectrum of CVD SLG shows a broad and asymmetric optical absorption band at 4.6 eV, which significantly red-shift from theoretic value (5.1 eV) of a symmetric absorption peak arising from the inter-band transitions in graphene.[36] According to recent calculation, the red shift is owing to strong resonant exciton effect raised in two-dimensional semimetals.[36] The transmission at low photon energies (0.6 eV < $E$ < 2 eV) exhibits a nearly flat absorption spectrum with a frequency independent absorbance of 2.3%, which results from linear dispersion in low energy bands of SLG.[37] Regarding to absorption spectroscopy of CVD BLG, it shows similar excitonic effects as those in CVD SLG, but with a less asymmetric optical absorption band at 4.6 eV. The 0.5 eV red-shift of absorption peak is close to the predicted value of 0.45 eV in AB stacked BLG.[36] Note that at low excitation energies, as shown in the inset of Figure 6, the absorbance of CVD BLG does not persist a expected constant value, for example 2x2.3% = 4.6%, but increases from 3.8% to 6% with increasing photon energy. Such a deviation from idealized flat optical absorbance should originate from the splitting of electronic bands in CVD BLG. According to inter-bands transitions between the split electronic bands around band edges, an increased absorbance in NIR range with increasing excitation energy has been demonstrated in recent calculation of NIR absorption in both AB and AA stacked BLG.[5]

**2.3 Carrier dynamics in strongly coupled BLG**

The electronic band structures play an important role in carrier dynamics, therefore the splitting of the electronic bands in CVD BLG, if there are any, would cause difference in carrier dynamics in CVD BLG and CVD SLG. The carrier dynamics of CVD BLG was investigated by transient absorption ($\Delta A/A$) spectroscopy in comparison to those of stacked layers of CVD SLG. Our recent study demonstrated that the stacked CVD SLG layers show layer number independent carrier dynamics and persisted nearly the same carrier dynamics as those in SLG due to weak layer interaction.[38] In this spectroscopy measurement, the CVD BLG and the stacked CVD SLG in two layers on quartz substrates were pumped with photon energy of 3.55 eV (350 nm) and probed with probe photon energy ($E_{pr}$) from 1.77 to 2.48 eV (500-700 nm). Figure 7 displays the transient absorption spectra of CVD BLG and stacked CVD SLG layers probed at 2.17 eV (570 nm). According to our previous study, the deconvolution-fit processes of kinetic curves resulted in two time components: $\tau_{rise}$ within 60 fs, a characteristic ultra-short time limited by our instrument resolution and $\tau_{decay}$ in a 100 fs time scale, a carrier relaxation time in agree with those measured in epitaxial graphene and graphite.[38-40] The $\tau_{rise}$, a time when the maximum bleaching happens at the probed energy level, originates from intra-band hot carrier thermalization governed by the carrier-carrier scattering.[38,40] Subsequently, these hot carriers relax their energies mainly by carrier-optical phonon scattering in the decay time of $\tau_{decay}$.[38-40] Note that the $\tau_{decay}$ of CVD BLG is 200 fs, larger than 130 fs of the stacked two layer CVD SLG. This slower decay time in CVD BLG should originate from the electron band splitting, which will be discussed in details as below.

Figure 8 shows dependence of the decay times versus probe photon wavelength. The decay time, for both CVD BLG and stacked two layer CVD SLG, has a increasing tendency with increasing probed photon wavelength, i.e. electrons at lower excited level live longer compared with the ones at higher energy level, analogous to those in graphite.[41] It is clear that the decay time in CVD BLG are always slower than those in stacked two layer CVD SLG and their discrepancy become larger at lower probed photon energy. Such phenomenon should be caused by slow relaxation time during the inter-band scattering at the edges of the split electron bands in BLG, which was previously demonstrated in the carriers dynamics of single walled carbon nanotube with split band structures.[42] The inset in Figure 8 indicates the split band structures in strongly coupled BLG and the pump/probe related optical transitions. After photo-excitation and rapid thermalization, the relaxation of carriers occurs mainly through intra-band carrier-phonon scatterings in each subbands down to their band-edge states on a 100 fs time scale, since the intra-band scattering rates are much higher than inter-band scattering rates.[40] Eventually, the carriers in upper subband edge will decay to the lower subband by inter-band scatterings, which are in picosecond time scale as observed in graphite and single wall carbon nanotube.[40,42,43] Such a slow decay time at the upper subband edge will accumulate carriers and slow down the intra-band decay in the upper subband according to the density of states effects and Pauli blocking.[40] Obviously, such slow down effect will become more apparent with decreasing energy level, in agreement with our observations. Therefore, consistent with the Raman spectroscopy and NIR absorption spectroscopy investigation, the carrier dynamics study also implies that electronic bands in CVD BLG should be in split form.

## 3. Conclusions

BLG was synthesized homogeneously in large area on Cu foil by thermal CVD. The growth mechanism of second graphene layer is found to be related to purity controlled catalystic activity of Cu surface. The 2D Raman band of CVD BLG splits into four components, indicating splitting of electronic bands due to strong interlayer coupling in CVD BLG. The splitting of electronic bands in CVD BLG is further evidenced by the study of NIR absorption spectroscopy, where interband transitions between the split electronic bands break down the frequency independent absorption found in CVD SLG. The ultraviolet photoemission spectroscopy invesigation also indiates CVD BLG possesses different electronic band structures from those of CVD SLG. The carrier decay times in CVD BLG probed by transient absorption spectroscopy are obviously slower than those of CVD SLG, which is is attributed to the influence of slow inter-band decay in edges of the split electronic bands. Our work shows that strongly coupled or even AB stacked BLG can be grown on Cu foils in large scale, which is of particular importance for special electronic and photonic device applications based on their split electronic bands.

**4. Experimental**

Graphene was grown on 34 um thick Cu foil (99.95% purity) by low pressure CVD. The Cu foils were heated up to 1000°C and then annealed for 30 min in flowing 10 sccm $H_2$ at 1 Torr. During CVD process, gas mixture of $H_2$ and $CH_4$ was flowed at 4 Torr with a rate of 10 sccm and 30 sccm, respectively. It takes 30 min for growth of a SLG and another 180 min for growth of a second layer on the first layer graphene. Subsequently, the samples were cooled down to room temperature (~20°C/min) with flowing $H_2$ under pressure of 1 Torr. With $Fe(NO_3)_3$ (0.2M) as Cu etchant, graphene on Cu foils were transferred onto Si wafer substrates with 285 nm $SiO_2$ cap layer for Raman spectroscopy and contrast spectroscopy characterizations. The Raman and contrast spectroscopies were performed on a WITEC CRM200 Raman system using a 100x objective lens with a numerical aperture (NA) of 0.95. The excitation source for Raman spectroscopy was a 532 nm laser (2.33 eV) with a laser power below 1 mW to avoid laser-induced heating. Illumination source for the contrast spectroscopy was tungsten halogen lamp (excitation range from 350 nm to 850 nm). For Raman imaging, the sample was placed on an X-Y piezo-stage and scanned under the illumination of laser light with a step size of 250 nm. UPS investigation of the as-grown graphene on Cu foils were performed on ESCALAB 250 (Thermo VG Scientific) with a He lamp as UV excitation. After transfer of graphene layers onto quartz substrates, UV-Vis-NIR absorption spectroscopy characteriztion was performed on LAMBDA 950 UV/Vis/NIR spectrophotometer (PerkinElmer, US) in 0.6-5 eV. The pump-probe measurements were performed in Biofemtolab (NTU). The output of titanium-sapphire (Legend Elite, Coherent) regenerative amplifier seeded by an oscillator (Micra, Coherent) was used as a pulse laser source: wavelength 800 nm, pulse width 65 fs, pulse repetition rate 1 kHz, and average power 3.5 W. The main part, 90%, of the radiation was converted into the UV (350 nm) by use of optical parametric oscillator (Topas, Light Conversion) with following second- and fourthharmonic generation that was used as pump pulse. The remaining 10% was used to generate white light continuum in $CaF_2$ plate, i.e. probe pulse.

**Acknowledgements:**

Yu Ting acknowledges the support by Singapore National Research Foundation under NRF RF Award No. NRFRF2010-07 and MOE Tier 2 MOE2009-T2-1-037.

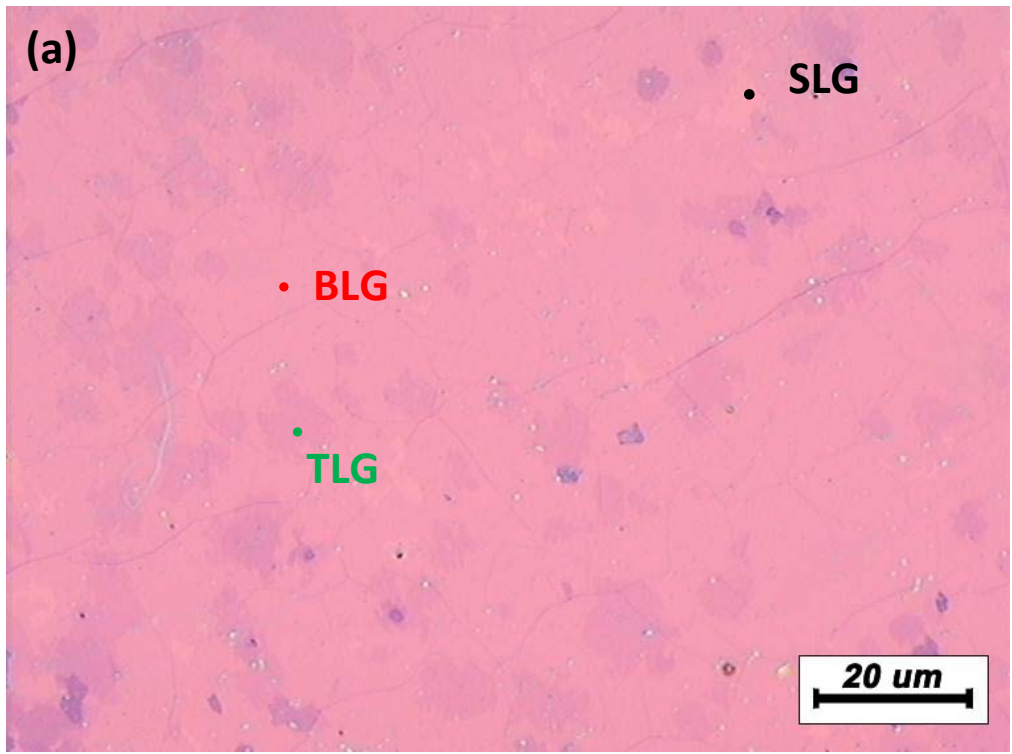
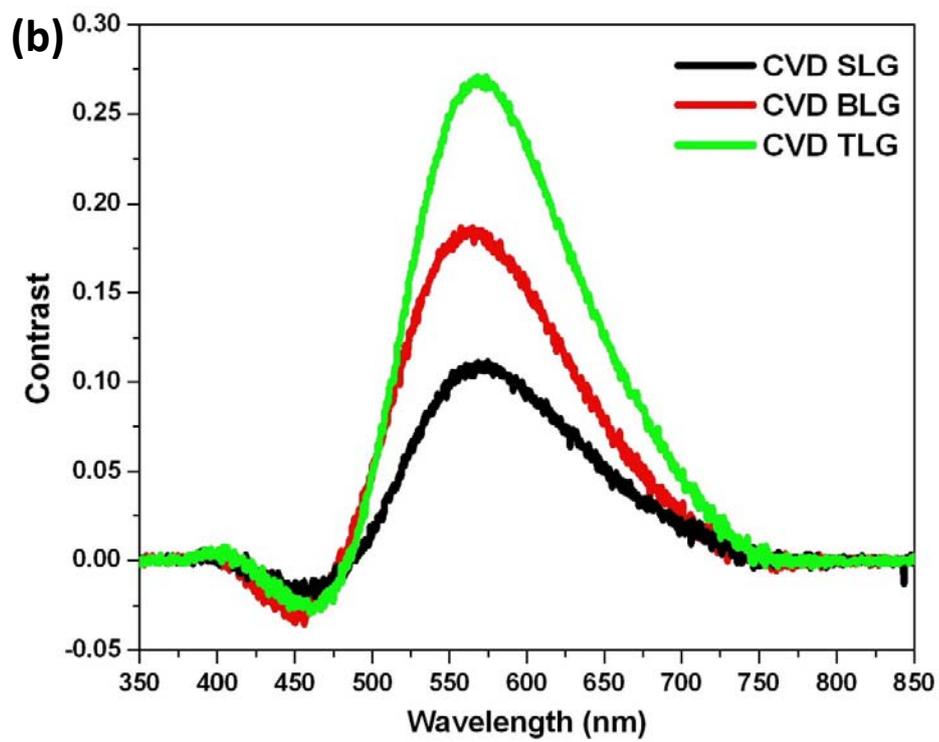

**Figure 1.** (a) Optical image of CVD bi-layer graphene sheet transferred onto the 285nm $SiO_2$/Si substrate. (b) Contrast spectra of CVD bi-layer graphene sheet on 285 nm $SiO_2$/Si substrate.

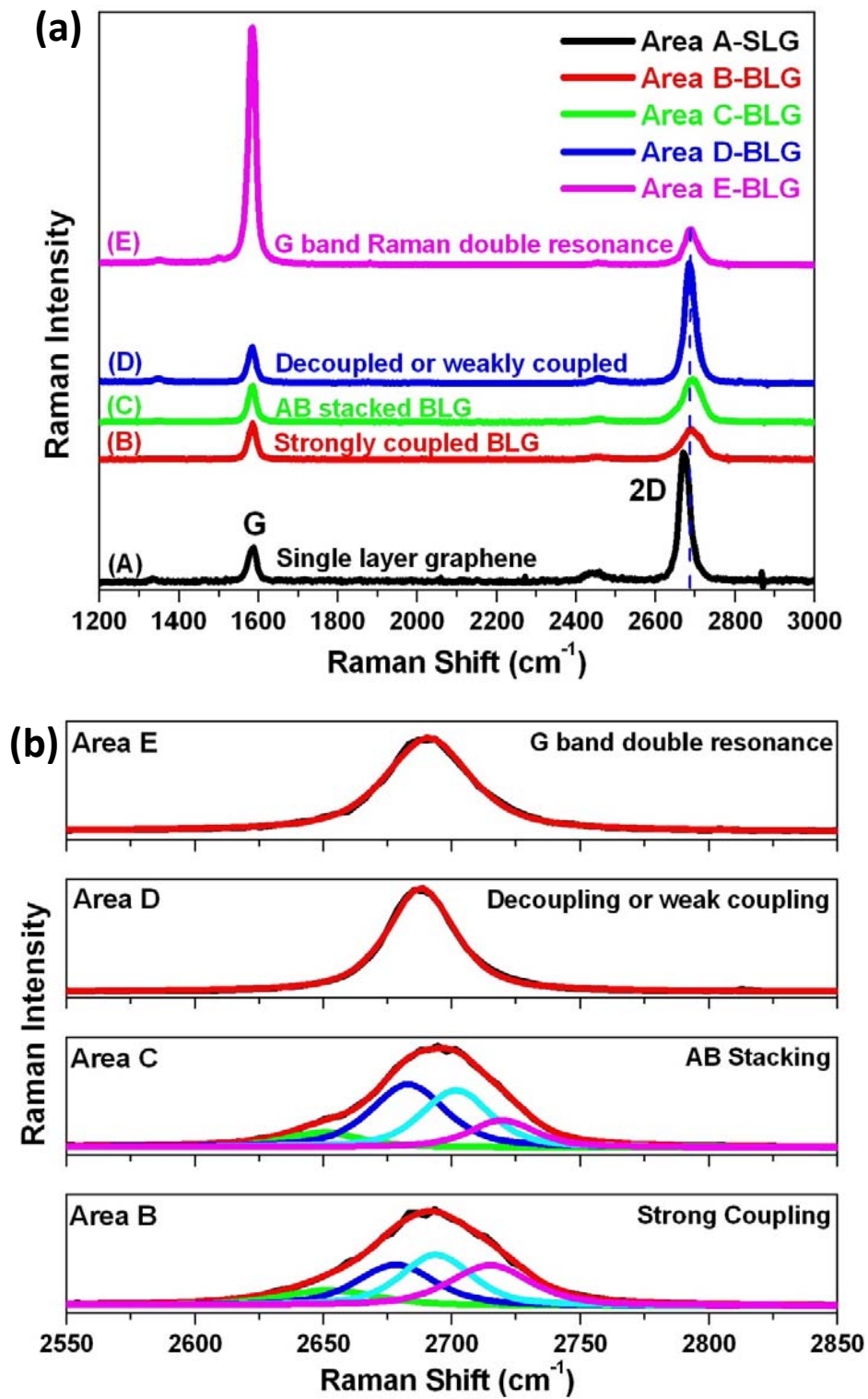

**Figure 2.** (a) Raman spectra recorded in CVD bi-layer graphene sheet. (b) Fittings of the 2D bands in the Raman spectra of CVD bi-layer graphene sheet.

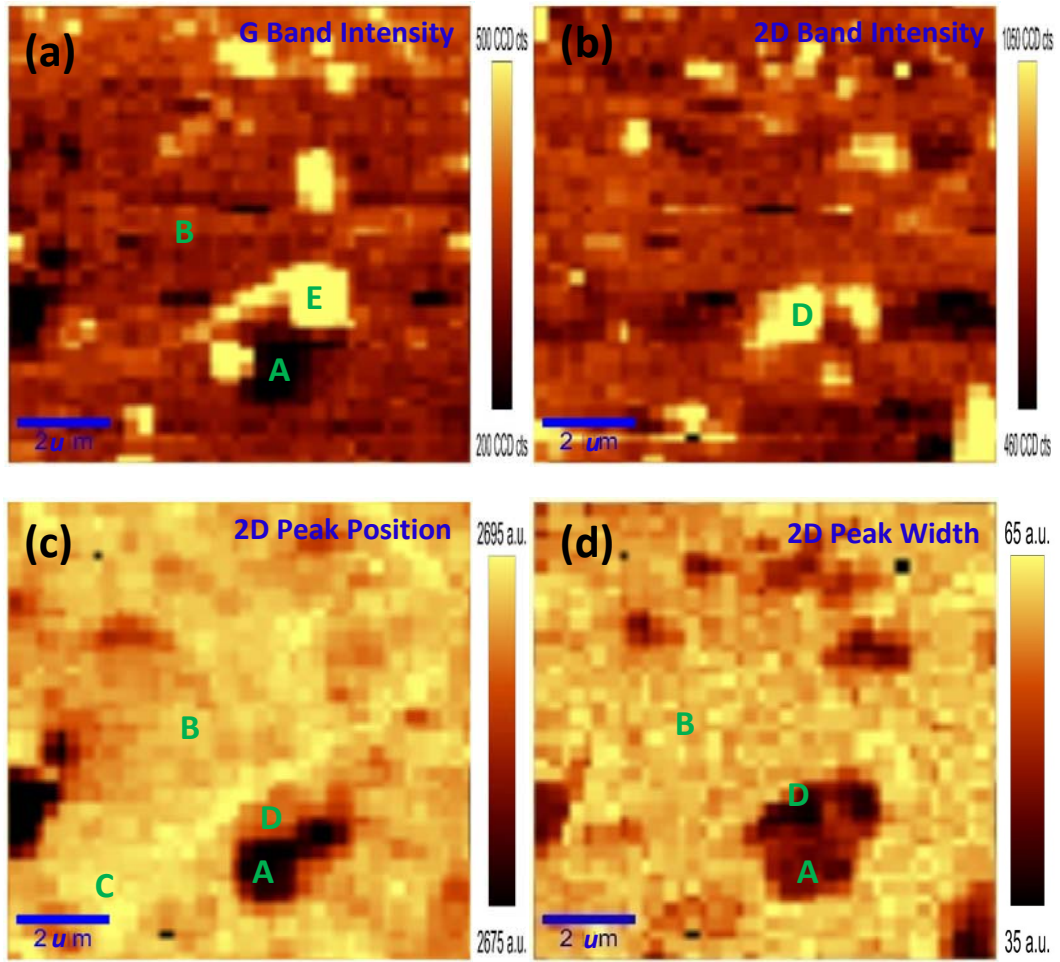

**Figure 3.** Raman imaging of CVD bi-layer graphene sheet. (a) G band intensity; (b) 2D band intensity; (c) 2D band peak position; (d) 2D band width.

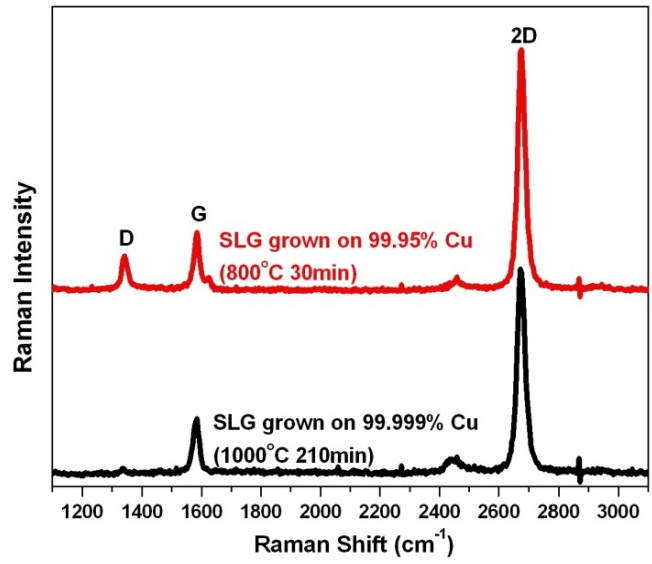

**Figure 4.** Raman spectra of CVD graphene grown on 99.999% Cu foil (1000C, 210 min) and 99.95% Cu foil (800C, 30 min).

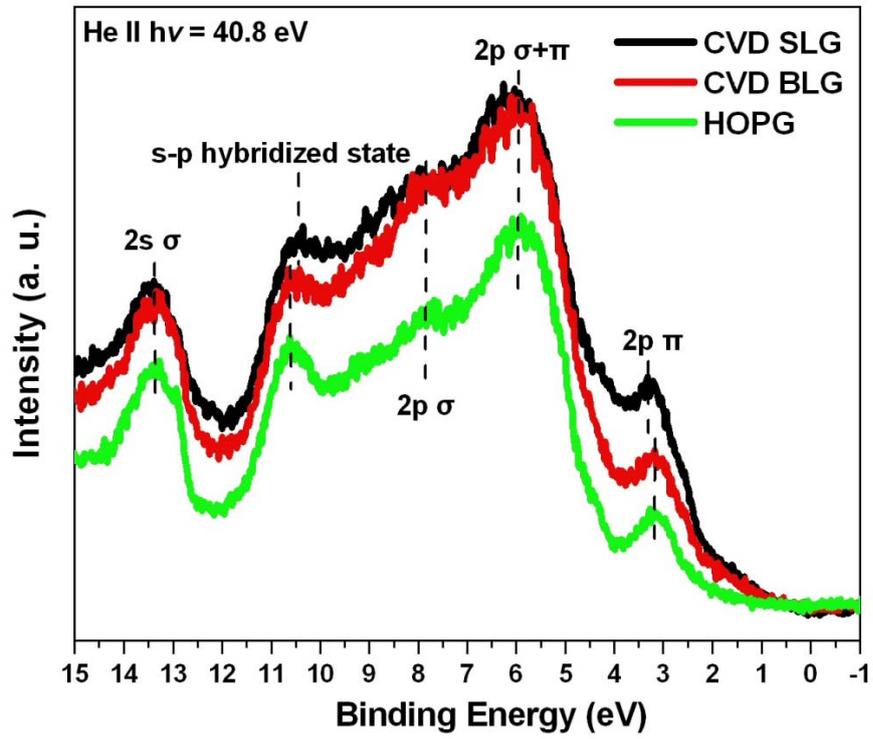

**Figure 5.** Relative variation of He II UPS valence band spectra of CVD single layer graphene, CVD bi-layer graphene, and HOPG.

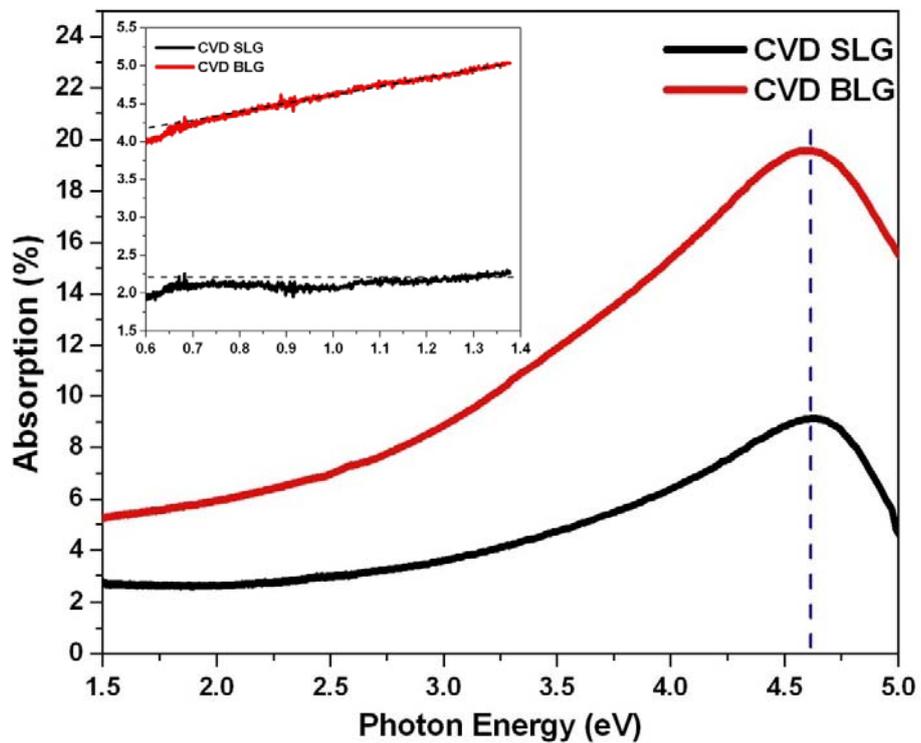

**Figure 6.** UV-Vis absorption of CVD single layer graphene and CVD bi-layer graphene. The inset shows their NIR absorption.

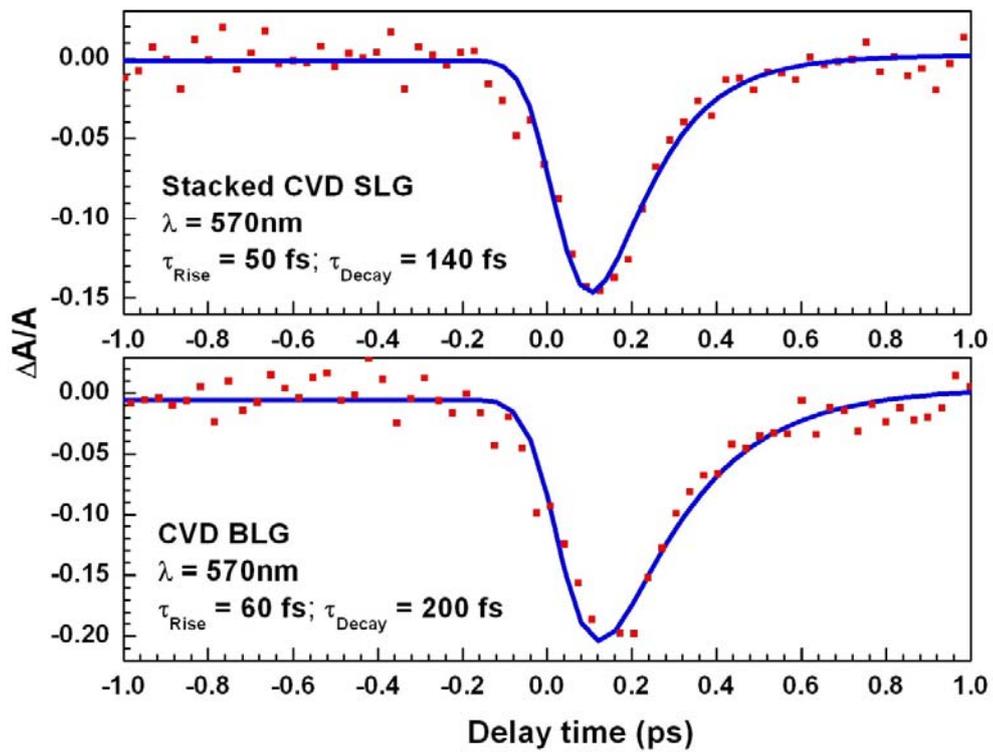

**Figure 7.** Transient absorption spectra ($\lambda_{excitation}$ = 350 nm; $\lambda_{probe}$ = 570 nm) of CVD bi-layer graphene and stacked two CVD single layer graphene films on quartz.

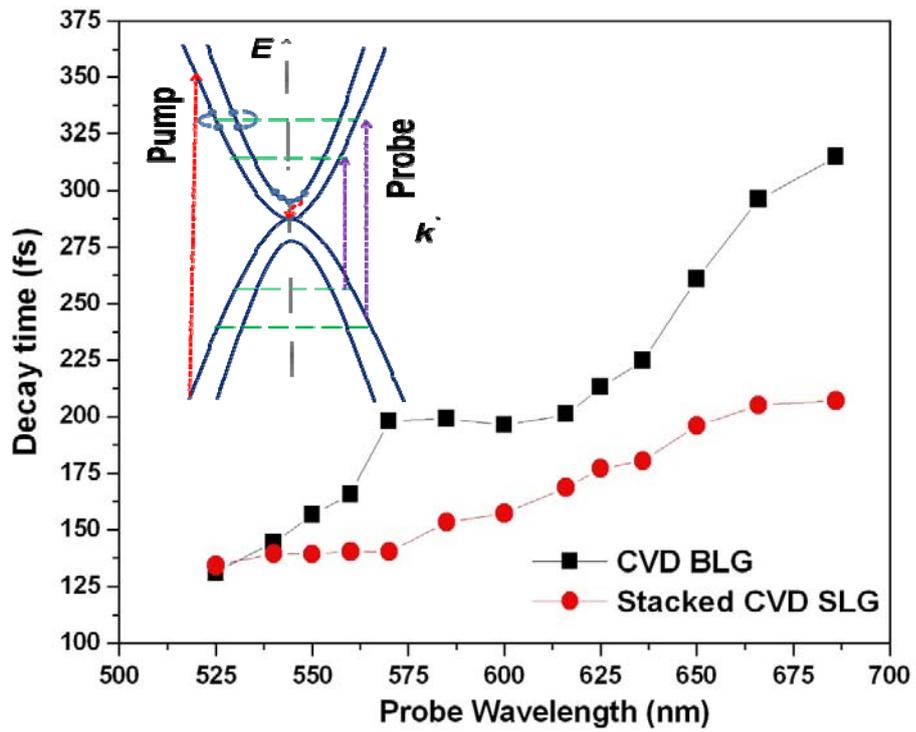

**Figure 8.** Decay times versus the probe photon wavelength in the range of 500-700 nm. The inset shows the pump/probe configuration in the electronic bands of bi-layer graphene and the interband/intraband decay process.